\documentclass{WileyMSP-template}

\usepackage{textcomp}
\usepackage{multicol}

\usepackage{graphicx}   
\usepackage{float}      
\usepackage{subcaption} 
\usepackage{adjustbox}  

\usepackage{tikz}       
\usepackage{standalone} 
\usepackage[siunitx]{circuitikz} 

\usepackage{amsmath}    
\usepackage{amssymb}    
\usepackage[version=3]{mhchem}     

\usepackage{hyperref}   
\usepackage{cleveref}   

\usepackage{xcolor}     
\usepackage{multirow} 
\usepackage{rotating}

\newcommand{\up}{\mathrm}
\newcommand{\s}{\;\!}

\begin{document}

\pagestyle{fancy}
\setlength{\headheight}{25pt} 

\setlength{\parindent}{0em}


\title{{Estimating Reaction Rate Constants from Impedance Spectra: Simulating the Multistep Oxygen Evolution Reaction}}

\maketitle


\author{Freja Vandeputte*}
\author{Bart van den Boorn*}
\author{Matthijs van Berkel} \\
\author{Anja Bieberle-Hütter} 
\author{Gerd Vandersteen}
\author{John lataire}

\begin{abstract}
The efficiency of water electrolysis in a photoelectrochemical cell is largely limited by the oxygen evolution reaction (OER) at its semiconductor photoanode.
Reaction rate constants are key to investigating the slow kinetics of the multistep OER, as they indicate the rate-determining step.
While these rate constants are usually calculated based on first-principles simulations, 
this research aims to estimate them from experimental electrochemical impedance spectroscopy (EIS) data.
Starting from a microkinetic model for charge transfer at the semiconductor-electrolyte interface,
an expression for the impedance as a function of the rate constants is derived. 
At lower potentials, the order of this impedance model is reduced, 
thus eliminating the rate constants corresponding to the last reaction steps.
Moreover, it is shown that EIS data from at least two potentials needs to be combined in order to uniquely identify the rate constants of a particular reduced order model. 
Therefore, this work details a sample maximum likelihood estimator that integrates not only multiple frequencies, but also multiple potentials simultaneously.
Measuring multiple periods of the current density and potential signals, allows this frequency domain estimator to take measurement uncertainty into account.
In addition, due to the large numerical range of the rate constants, various scaling methods are implemented to achieve numerical stability.
To find suitable initial values for the highly nonlinear optimization problem, different global estimation methods are compared.
The complete estimation procedure of the rate constants is illustrated on simulated EIS data of a hematite photoanode.

\end{abstract}


\keywords{rate constants, oxygen evolution reaction, photoelectrochemical cell, electrochemical impedance spectroscopy, frequency domain identification, maximum likelihood estimation}



\begin{affiliations}
F. Vandeputte*, G. Vandersteen, J. Lataire\\
ELEC - Department of Electricity\\
Vrije Universiteit Brussel\\
Pleinlaan 2, 1050 Brussels, Belgium\\
E-mail: Freja.Vandeputte@vub.be
\end{affiliations}

\begin{affiliations}
B. van den Boorn*, A. Bieberle-Hütter\\
Electrochemical Materials and Interfaces\\
DIFFER - Dutch Institute for Fundamental Energy Research\\
De Zaale 20, 5612 AJ Eindhoven, The Netherlands\\
E-mail: B.F.H.vandenBoorn@differ.nl
\end{affiliations}

\begin{affiliations}
M. van Berkel\\
Energy Systems and Control\\
DIFFER - Dutch Institute for Fundamental Energy Research\\
De Zaale 20, 5612 AJ Eindhoven, The Netherlands\\
Control Systems, Department of Electrical Engineering \\
Eindhoven University of Technology \\
Flux, Groene Loper 19, 5612 AP Eindhoven, The Netherlands
\end{affiliations}

\pagebreak
\begin{multicols}{2}
\setlength{\parindent}{1em}





\section{Introduction}
Hydrogen generation by the electrolysis of water using solar energy is a promising option for achieving clean and sustainable energy production.
This solar water-splitting can be done using a photoelectrochemical cell (PEC).
Unfortunately, the current efficiencies of \textit{green} hydrogen production by PEC water-splitting cannot compete with those of \textit{grey} hydrogen production from fossil fuels, 
nor with the solar-to-electricity efficiencies in photovoltaic cells. 
To be able to improve the solar-to-hydrogen conversion efficiencies in PECs, it is important to characterise the limiting processes. \\[-1mm]

A PEC typically consists of a semiconductor photoanode and a metallic cathode, separated by a liquid electrolyte \cite{krol2012photoelectrochemical}.
When the photoanode is illuminated, it generates electron-hole pairs, which in turn drive the water electrolysis. 
At the photoanode-electrolyte interface, the holes oxidise water to form oxygen, called the oxygen evolution reaction (OER),
while at the cathode, the electrons reduce water to make hydrogen, known as the hydrogen evolution reaction (HER).
Of these two half-reactions, the OER is the performance-limiting process \cite{jiang2017}.
This is amongst others because the OER requires the transfer of four carriers to form one $\up{O}_2$ molecule, 
whereas the HER involves only two carrier transfers to create an $\up{H}_2$ molecule. \\[-1mm]

We consider the reaction mechanism of the OER in which the four carrier transfers correspond to different reaction steps, resulting in different intermediate species \cite{rossmeisl2007}.
Each reaction step of the multistep OER is characterised by a forward and a backward reaction rate.
These reaction rate constants are vital for understanding the OER kinetics, 
since they can indicate the rate-determining step (RDS). 
Furthermore, they can be used to predict electrochemical data and to compare kinetics of different photoanode materials.
Different methods for determining rate constants have been proposed in the literature. 
Traditionally, most studies use first-principles density functional theory (DFT) 
to simulate the Gibbs free energies of the intermediate species, from which the rate constants are then calculated 
\cite{norskov2004origin}\cite{zhang2016oxygen}\cite{ping2017reaction}\cite{george2019}.
However, DFT is computationally expensive and assumes idealised configurations at temperatures near zero Kelvin.
Alternatively, other works identify the reaction rate constants from experimental data, for example by fitting a theoretical current density model to a linear sweep voltammetry (LSV) plot \cite{alobaid2018mechanism} or Tafel plot \cite{zhang2020advances}. \\[-1mm]

This work aims to estimate the reaction rate constants of the multistep OER from experimental electrochemical impedance spectroscopy (EIS) data.
\textit{Garcia et al. (2017)} \cite{garcia2017kinetic} also determined OER rate constants by fitting EIS spectra.
However, while they used a least squares estimator to fit their model, this study uses the sample maximum likelihood (SML) estimator. 
Measuring multiple periods of the current density and potential signals, allows to weigh with the measurement uncertainty. 
This makes the SML estimate consistent, meaning that it converges to the true values of the rate constants \cite{Pintelon2012}.
We will show that the rate constants are uniquely identifiable if EIS data from multiple applied potentials is combined.
Therefore, we implemented a novel SML estimator that integrates multiple frequencies and potentials simultaneously. 
Additionally, the large numerical range of the rate constants required various scaling methods to improve numerical stability.\\[-1mm]

Moreover, the aforementioned works \cite{alobaid2018mechanism},\cite{zhang2020advances}, and \cite{garcia2017kinetic} all model both the forward and the backward rates as an exponential function of the overpotential and the charge transfer coefficient. 
In \cite{garcia2017kinetic}, the pre-exponential rate constant and the charge transfer coefficient even change with the applied potential.
Although this expression can be empirically substantiated for metal electrodes, it makes less sense for semiconductor photoanodes, 
where charge transfer happens primarily via the valence or conduction bands \cite{book2015}.
Hence, this work uses the rate constant expressions from Gerischer theory \cite{gerischer1969charge}, where the pre-exponential rate constants are potential-independent and therefore truly constant. \\[-1mm]

The remainder of this paper is organised as follows. 
First, in Section 2, we derive the impedance model as a function of the rate constants.
Next, in Section 3, we explain how, at lower potentials, the order of this model can be reduced.
Section 4 discusses the gathering and preprocessing of the frequency domain EIS data.
The sample maximum likelihood estimation and the global optimization algorithms used for the initial estimation are revisited in section 5.
Section 6 investigates the identifiability of the rate constants, both in a given frequency band and over multiple potentials.
Finally, Section 7 contains simulation results of a hematite photoanode.


\section{Model}
This section shows how the impedance at the photoanode-electrolyte interface 
can be modelled as a function of the rate constants.
Similar to \textit{George et al. (2019)} \cite{george2019}, we start from the reaction steps of the multistep OER, 
and linearise the corresponding rate equations and current density equation around an equilibrium point. 
The impedance is then derived from the resulting linear state-space model. 
In this work, however, we also eliminated the unmeasurable equilibrium species coverages 
from the model, by writing them in terms of the rate constants. 

\subsection{Microkinetic model}
\label{sec:MicroModel}
The OER in an alkaline environment is given by 
\begin{equation}
   \ce{4OH- + 4h+ -> O2 + 2H2O}
\end{equation}
which can be partitioned into four single-hole transfer reactions \cite{rossmeisl2007}
\begin{align}
   \cee{{*} + OH- + h+ & <=>[\tilde{K}_{f1}][\tilde{K}_{b1}] OH^*} \label{eq:reac1}\\
   \cee{OH^* + OH- + h+ & <=>[\tilde{K}_{f2}][\tilde{K}_{b2}] O^* + H2O} \label{eq:reac2}\\
   \cee{O^* + OH- + h+ & <=>[\tilde{K}_{f3}][\tilde{K}_{b3}] OOH^*} \label{eq:reac3}\\
   \cee{OOH^* + OH- + h+ & <=>[\tilde{K}_{f4}][\tilde{K}_{b4}] O^*2 + H2O} \label{eq:reac4}
\end{align}
and an oxygen desorption reaction
\begin{equation}
   \ce{O^*2 ->[K_{f5}] O2 + {*}}
   \label{eq:reac5}
\end{equation}
where $\up{OH}^*,\up{O}^*,\up{OOH}^*$ and $\up{O}^*_2$ are the intermediate species adsorbed on the photoanode surface, and $*$ represents the adsorption site. 
The forward and backward reaction rates of the hole transfer reactions are respectively denoted as $\tilde{K}_{fi}$ and $\tilde{K}_{bi}$ for $i=1,\dots,4$. 
These reaction rates, expressed in cm/s, 
are related to the rate constants $k_{fi}$ and $k_{bi}$, expressed in cm$^4$/s,
as obtained from Gerischer theory for hole transfer via the valence band of the semiconductor, as follows \cite{book2015}
\begin{align}
   \tilde{K}_{fi} &= k_{fi}\s p_s\\
   \tilde{K}_{bi} &= k_{bi}\s N_V
\end{align}
Here, $p_s$, in cm$^{-3}$, is the hole density at the surface of the semiconductor,
and $N_V$, in cm$^{-3}$, is the state density in the valence band.
While the latter is a constant, the former increases exponentially with the potential $u$ in V as
\begin{equation}
   p_s = p_{s0} \up{exp} \bigg(\frac{e\s u}{k_BT}\bigg)
\end{equation}
with  $p_{s0}$, in cm$^{-3}$, the hole density at the surface in the dark at equilibrium,
$e$ the elementary charge, $k_B$ the Boltzmann constant and $T$ the temperature.
The reaction rate of the desorption reaction is denoted as $K_{f5}$ and is assumed to be constant.  

\subsection{Nonlinear state-space model}
\label{sec:NLSS}
Denoting $\theta_i \in [0,1]$ as the fractional coverages of the intermediate species $i$, 
for $i = \up{OH}^*,\up{O}^*,\up{OOH}^*$ and $\up{O}_2^*$,
and $\theta_* \in [0,1]$ as the fraction of free adsorption sites,
the rate equations of the hole transfer reactions \eqref{eq:reac1}-\eqref{eq:reac5} are formulated as
\begin{align}
   \dot \theta_{\up{OH}} &= (\tilde{K}_{f1}\s\theta_{*}\s x_{\up{OH}} - \tilde{K}_{b1}\s\theta_{\up{OH}}) \nonumber\\
   &- (\tilde{K}_{f2}\s\theta_{\up{OH}}\s x_{\up{OH}} - \tilde{K}_{b2}\s\theta_{\up{O}}\s x_{\up{H_2O}}) \label{eq:rateeq1}\\
   \dot \theta_{\up{O}} &= (\tilde{K}_{f2}\s\theta_{\up{OH}}\s x_{\up{OH}} - \tilde{K}_{b2}\s\theta_{\up{O}}\s x_{\up{H_2O}}) \nonumber\\
   &- (\tilde{K}_{f3}\s\theta_{\up{O}}\s x_{\up{OH}} - \tilde{K}_{b3}\s\theta_{\up{OOH}}) \label{eq:rateeq2}\\
   \dot \theta_{\up{OOH}} &= (\tilde{K}_{f3}\s\theta_{\up{O}}\s x_{\up{OH}} - \tilde{K}_{b3}\s\theta_{\up{OOH}}) \nonumber\\
   &- (\tilde{K}_{f4}\s\theta_{\up{OOH}}\s x_{\up{OH}} - \tilde{K}_{b4}\s\theta_{\up{O_2}}\s x_{\up{H_2O}}) \label{eq:rateeq3}\\
   \dot \theta_{\up{O_2}} &= (\tilde{K}_{f4}\s\theta_{\up{OOH}}\s x_{\up{OH}} - \tilde{K}_{b4}\s\theta_{\up{O_2}}\s x_{\up{H_2O}}) \nonumber\\
   &- ({K}_{f5}\s\theta_{\up{O_2}}) \label{eq:rateeq4}
\end{align}
with
\begin{equation}
   \theta_{*} = 1-\theta_{\up{OH}}-\theta_{\up{O}}-\theta_{\up{OOH}}-\theta_{\up{O_2}}
   \label{eq:balanceeq}
\end{equation}
The current density $j$ in A/cm$^2$ is then given by 
\begin{align}
   j = e&N_0(\tilde{K}_{f1}\s\theta_{*}\s x_{\up{OH}} - \tilde{K}_{b1}\s\theta_{\up{OH}} \nonumber\\ 
   &+ \tilde{K}_{f2}\s\theta_{\up{OH}}\s x_{\up{OH}} - \tilde{K}_{b2}\s\theta_{\up{O}}\s x_{\up{H_2O}} \nonumber\\ 
   &+ \tilde{K}_{f3}\s\theta_{\up{O}}\s x_{\up{OH}} - \tilde{K}_{b3}\s\theta_{\up{OOH}} \nonumber\\ 
   &+ \tilde{K}_{f4}\s\theta_{\up{OOH}\s }x_{\up{OH}} - \tilde{K}_{b4}\s\theta_{\up{O_2}}\s x_{\up{H_2O}})
   \label{eq:current}
\end{align}
with $N_0$ the density of adsorption sites on the electrode surface in cm$^{-3}$. \\[-1mm]

The terms $x_{\up{OH}}$ and $x_{\up{H_2O}}$ are the mole fractions of hydroxide ions and water in the electrolyte respectively. They are calculated from the electrolytes pH. 
Since $x_{\up{OH}}$ appears exclusively in combination with the forward rates $\tilde{K}_{fi}$, and $x_{\up{H_2O}}$ only with the backward rates $\tilde{K}_{b2}$ and $\tilde{K}_{b4}$, we introduce the scaled reaction rates
\begin{align}
   {K}_{fi} &= \tilde{K}_{fi}\s x_{\up{OH}} \qquad \text{for } i = 1,...,4 \\
   {K}_{bi} &= \begin{cases} \tilde{K}_{bi}  & \text{for } i = 1,3 \\ \tilde{K}_{bi}\s x_{\up{H_2O}} & \text{for } i = 2,4 \end{cases}
\end{align}
Note that the reaction rates in equations \eqref{eq:rateeq1}-\eqref{eq:rateeq4} are actually local rates expressed in 1/s.  
However, assuming that they are constant in the direction perpendicular to the electrode surface, these local rates in 1/s are equal to the rates in cm/s divided by the unit length of 1 cm. 

\subsection{Linearised state-space model}
\label{sec:LSS}
The rate equations \eqref{eq:rateeq1}-\eqref{eq:rateeq4} and the current density equation \eqref{eq:current} describe a nonlinear state-space model with the potential $u$ as input, the current density $j$ as output and the fractional coverages of the intermediate species as the four states, collected in the state vector
\begin{equation}
   {\bf\Theta} = \begin{bmatrix} \theta_{\up{OH}} & \theta_{\up{O}} & \theta_{\up{OOH}} & \theta_{\up{O_2}} \end{bmatrix}^\top
\end{equation}
The model is nonlinear since the forward reaction rates depend exponentially on the input potential $u$.
Linearising the model around a given operating point $({\bf\Theta}_{\mathrm{eq}},u_{\mathrm{eq}})$, with $u_{\mathrm{eq}}$ the equilibrium potential and ${\bf\Theta}_{\mathrm{eq}}$ the vector of equilibrium species coverages at this potential, gives the linear state-space model \cite{george2019}
\begin{align}
   \dot {\tilde{\bf\Theta}} &= A {\tilde{\bf\Theta}} + B\tilde{u} \label{eq:stateeq}\\
   \tilde{j} &= C {\tilde{\bf\Theta}} + D\tilde{u} \label{eq:outputeq}
\end{align}
where the tilde denotes a perturbation around the operating point
\begin{align}
   {{\bf\Theta}} &= {{\bf\Theta}_{\mathrm{eq}}} + {\tilde{\bf\Theta}} \\
   u &= u_\mathrm{eq} + \tilde u \\
   j &= j_\mathrm{eq} + \tilde j
\end{align}
and the state-space matrices are obtained as
\begin{equation}
   A \hspace{-0.8mm}=\hspace{-1.5mm} \begin{bmatrix}
   \!-\!K_{f1}\!+\!a_{11} & \!-\!K_{f1}\!+\!K_{b2} & \!-\!K_{f1} & \!-\!K_{f1} \\
   K_{f2} & a_{22} & K_{b3} & 0 \\
   0 & K_{f3} & a_{33} & K_{b4} \\
   0 & 0 & K_{f4} & a_{44}
   \end{bmatrix}
   \label{eq:Amatrix}
\end{equation}
with
\begin{align}
   a_{11} &= -K_{b1}-K_{f2} \\
   a_{22} &= -K_{b2}-K_{f3} \\
   a_{33} &= -K_{b3}-K_{f4} \\
   a_{44} &= -K_{b4}-K_{f5}
\end{align}
and
\begin{equation}
   B = \frac{e}{k_BT}\begin{bmatrix}
   K_{f1}\s\theta_{\up{*}}\;\ -K_{f2}\s\theta_{\up{OH\phantom{O}}} \\
   K_{f2}\s\theta_{\up{OH}}-K_{f3}\s\theta_{\up{O\phantom{OO}}} \\
   K_{f3}\s\theta_{\up{O}}\;\; -K_{f4}\s\theta_{\up{OOH}} \\
   K_{f4}\s\theta_{\up{OOH}}
   \end{bmatrix}
   \label{eq:Bmatrix}
\end{equation}
\begin{equation}
   C = eN_0\begin{bmatrix}
   -K_{f1}-K_{b1}+K_{f2} \\
   -K_{f1}-K_{b2}+K_{f3} \\
   -K_{f1}-K_{b3}+K_{f4} \\
   -K_{f1}-K_{b4}
   \end{bmatrix}^\top
   \label{eq:Cmatrix}
\end{equation}
\begin{equation}
   D = \frac{e^2N_0}{k_BT}
   (K_{f1}\s\theta_{\up{*}}+K_{f2}\s\theta_{\up{OH}} 
   +K_{f3}\s\theta_{\up{O}}+K_{f4}\s\theta_{\up{OOH}})
   \label{eq:Dmatrix}
\end{equation}
Here, all forward rates $K_{fi}$ are evaluated in the equilibrium potential $u_{\mathrm{eq}}$ and all the coverages $\theta_i$ are the equilibrium species coverages at this potential.

\subsection{Equilibrium species coverages}
\label{sec:equilibrium-species-coverages}
Unlike the current density and the potential,
the fractional coverages of intermediate species are not measurable and thus unknown quantities.
However, their equilibrium values can be written as function of the reaction rates by expressing that per definition, they do not vary over time, i.e. $\dot{\bf\Theta}_{\mathrm{eq}} = 0$.
Hence, it follows from the rate equations \eqref{eq:rateeq1}-\eqref{eq:rateeq4} and from relation \eqref{eq:balanceeq} that 
\begin{equation}
   A{\bf\Theta}_{\mathrm{eq}} = -K_{f1}{\bf e}_1  
   \Rightarrow {\bf\Theta}_{\mathrm{eq}} = -K_{f1}A^{-1}{\bf e}_1
   \label{eq:eq_species_coverages}
\end{equation}
where the matrix $A$ and the rate $K_{f1}$ are evaluated in the potential $u_{\mathrm{eq}}$, and the vector ${\bf e}_1$ denotes the unit vector with the first element one and the others zero.

\subsection{Impedance model}
\label{sec:Zmodel}
The admittance $Y(s)$ corresponding to the linearised state-space model \eqref{eq:stateeq}-\eqref{eq:outputeq} is obtained as
\begin{equation}
   Y(s) = \frac{J(s)}{U(s)} =  C(sI_n-A)^{-1}B + D 
   \label{eq:admittance}
\end{equation}
where $U(s)$ and $J(s)$ are the Laplace transforms of the input potential and the output current density respectively, and $I_n$ denotes the identity matrix.
The impedance spectrum at the potential $u_{\up{eq}}$
is then simply the inverse of this admittance, i.e. $Z(s) = Y^{-1}(s)$.
The admittance \eqref{eq:admittance} can be written as a transfer function 
of the form
\begin{equation}
   Y(s) = C_0 \, \frac{B(s)}{A(s)}
   = C_0 \, \frac{b_ns^n + b_{n-1}s^{n-1} + \dots + b_0}{s^n + a_{n-1}s^{n-1} + \dots + a_0}
   \label{eq:TF}
\end{equation}
with the scaling factor
\begin{equation}
   C_0 = \dfrac{e^2N_0}{k_BT}
\end{equation}
Since the state-space model \eqref{eq:stateeq}-\eqref{eq:outputeq} has four states, related to the four intermediate species, this is a transfer function of order $n=4$.

\section{Reduced order model}
\label{sec:ReducedOrderModel}
%
At low potentials, only the first $n<4$ species are actually adsorbed on the surface, while the others are reacting away very fast, such that their coverages $\theta_i$ can be considered to be zero.
Therefore, these states can be eliminated and the state-space model \eqref{eq:stateeq}-\eqref{eq:outputeq} simplifies to 
\begin{align}
   {\dot {\tilde{\bf\Theta}}} _{[1:n]} &= A_{[1:n,1:n]}  {\tilde{\bf\Theta}}_{[1:n]} + B_{[1:n]}\tilde u 
   \label{eq:stateeqred}\\
   \tilde j &= C_{[1:n]}  {\tilde{\bf\Theta}}_{[1:n]} + D_n\tilde u
   \label{eq:outputeqred}
\end{align}
where the subscript $[1\!:\!n]$ refers to the first $n$ elements of the vector, 
the submatrix $A_{[1:n,1:n]}$ consists of the first $n$ rows and columns of the matrix $A$, 
and $D_n$ is obtained by taking the first $n+1$ terms of \eqref{eq:Dmatrix}.
Similarly, the equilibrium species coverages are obtained as 
\begin{equation}
   {\bf\Theta}_{{\mathrm{eq}},[1:n]} = -K_{f1}A_{[1:n,1:n]}^{-1}{\bf e}_1
\end{equation}
Hence, the fourth order transfer function reduces to an $n$th order transfer function \eqref{eq:TF} with $n<4$.
Note that this reduced order model only contains the rate constants of the first reaction steps, specifically $k_{fi}$ for $i = 1,\dots,n+1$ and $k_{bi}$ for $i = 1,\dots,n$. \\[-1mm]

\begin{figure} [H]
\centering
\includegraphics[scale = 0.75]{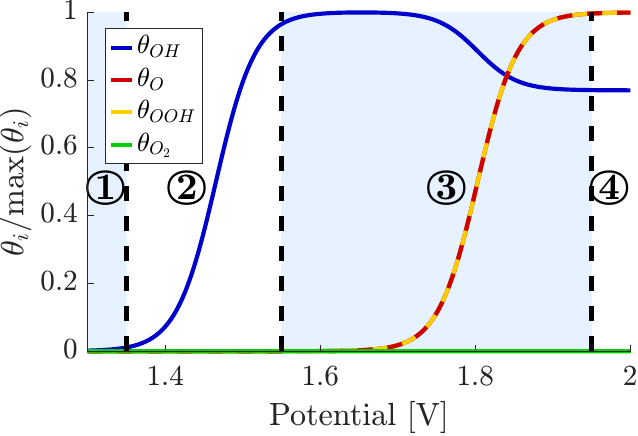}\hspace{5mm} %
\caption{Normalised equilibrium species coverages at different potentials for a hematite photoanode. The coloured regions correspond to different reduced model orders. The normalised coverage of O$_2^*$ only starts to increase at higher potentials. \vspace{-4mm}}
\label{fig:SpeciesCoverage}
\end{figure}

\begin{figure} [H] 
\centering
\includegraphics[scale = 0.75]{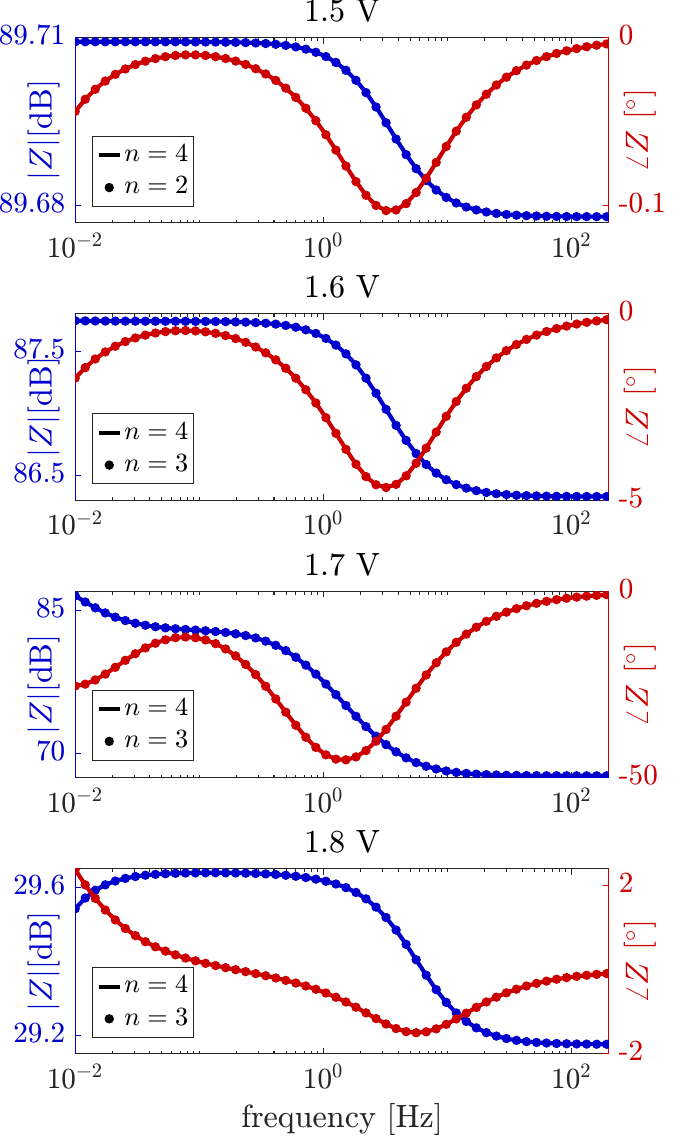} %
\caption{Bode plots of the fourth order model compared with the second order model at 1.5~V (top), and with the third order model at 1.6~V, 1.7~V and 1.8~V (bottom).\vspace{-4mm}}
\label{fig:BodePlotsReducedOrder}
\end{figure}

\textbf{Figure}~\ref{fig:SpeciesCoverage} shows the normalised equilibrium species coverages at different potentials, calculated using \eqref{eq:eq_species_coverages}, for the hematite-water interface simulated in Section~\ref{sec:Simulation}. 
Theoretically, the applied potential must be larger than the $\up{H_2O}/\up{O_2}$ redox potential of 1.23~V for the water oxidation to happen.
However, for potentials lower than 1.35~V, the coverage of OH$^*$ remains negligible, so the OER does not actually take place yet.
The few, very low-frequent charge transfer reactions that do occur, can be modelled by the first order model.
From 1.35~V to 1.55~V, the OH$^*$ coverage increases, and the fourth order model can be approximated by the second order model.
For potentials between 1.55~V and 1.95~V, the intermediate species O$^*$ and OOH$^*$ are adsorbed on the surface as well, and the third order model remains a good approximation for frequencies up to 1~MHz. 
This is illustrated in \textbf{Figure}~\ref{fig:BodePlotsReducedOrder}, which shows the Bode plots of the fourth order impedance model and the appropriate reduced order model at different potentials.
In all of the plots, the relative error between the models is less than $10^{-5}$ at all the frequencies in the considered frequency band of 10~mHz to 200~Hz. \\[-1mm] 

Remark that the impedance $Z(s)$ only models the charge transfer reactions at the photoanode-electrolyte interface.
Experimentally measured EIS data, however, also contains other contributions, such as resistive, capacitive or diffusive components.
Hence, to compare simulated and experimental EIS data, we need an equivalent circuit model (ECM).
A commonly used ECM consists of a series resistance and a bulk capacitance in parallel with the charge transfer impedance $Z(s)$ \cite{george2019}. 
The total impedance of this equivalent circuit $Z_\up{ECM}(s)$ will have a realistically measurable magnitude and phase.
In other words, it is not actually required to measure with a precision of 0.01~dB or 0.1$^\circ$, as the impedance plot at 1.5~V in {Figure}~\ref{fig:BodePlotsReducedOrder} suggests.
A more detailed discussion of integrating the charge transfer impedance in an ECM is given in the companion paper \cite{Bart2025}.

\section{Frequency domain data}
\label{sec:FreqDomainData}
As mentioned above, the impedance of our electrochemical system will be measured with a technique called electrochemical impedance spectroscopy. 
Specifically, this frequency domain, nonparametric estimation method calculates the impedance by dividing the discrete Fourier transform (DFT) spectra of measured voltage and current signals.
Potentiostatic EIS is performed by applying a potential excitation and measuring the current density response. 
The input potential consists of a small perturbation $u_{\up{ms}}(t)$ on top of the equilibrium potential $u_{\mathrm{eq}}$,
\begin{equation}
   u(t) = u_{\mathrm{eq}} + u_{\up{ms}}(t)
\end{equation}
The perturbation signal is an odd random phase multisine \cite{ORPEIS2009}
\begin{equation}
    u_{\up{ms}}(t) = \sum_{k\in\mathbb{K}_\mathrm{exc}} \alpha_{k} \sin{\left(\omega_k t+\varphi_{k}\right)}
    \label{eq:multisine}
\end{equation}
where $\alpha_{k}$ are user-defined amplitudes, 
$\omega_k = {2\pi k/T}$, with $T$ the period of the multisine, are the excited frequencies,
and $\varphi_{k}$ are random phases uniformly distributed in $[0,2\pi]$.
To avoid being corrupted by even nonlinear distortions in the output spectrum, the excited frequencies are chosen such that only odd harmonics are excited, i.e.\ $\mathbb{K}_\mathrm{exc}\subset2\mathbb{N}+1$. 
Additionally, the excited frequencies are quasi-logarithmically distributed, thus spanning a large frequency band of multiple decades. \\[-1mm]

The obtained multisine is scaled such that it has the desired root-mean-square (RMS) value.
If this RMS value is sufficiently small, the electrochemical system can be considered to be both linear and time-invariant, and thus the DFT spectra 
of the measured voltage and current signals, respectively denoted as $U(k)$ and $J(k)$, only have contributions at the excited frequencies.
Hence, the maximum likelihood estimation in Section~\ref{sec:MLE} will be restricted to the excited frequencies $\omega_k$ for $k\in \mathbb{K}_{\up{exc}}$.
Remark that the constant equilibrium potential $u_\mathrm{eq}$, and its corresponding equilibrium current density $j_\mathrm{eq}$, result in components at the zero frequency, but DC will not be used for the estimation.\\[-1mm]

We assume an errors-in-variables setup, where the measured signals are disturbed by additive zero-mean circular complex normally distributed noise \cite{Pintelon2012}. 
Measuring $M$ periods of the signals allows to reduce the noise level by averaging.
Denoting $U_m(k)$ and $J_m(k)$ for $m = 1,\dots,M$ as the DFT spectra of each period, the sample means are calculated as 
\begin{align}
   \hat X(k) &= \frac{1}{M}\sum_{m=1}^M X_m(k) \label{eq:meanX}
\end{align}
with $X = U,J$, and the sample noise covariances as 
\begin{align}
    \hat \sigma_{XY}^2(k) &= \frac{1}{M\!-\!1}\!\sum_{m=1}^{M} \!\!\left(\!X_m(k)\!-\!\hat X(k)\!\right)\!\!\!\left(\!Y_m(k)\!-\!\hat Y(k)\!\right)^{\!*} \label{eq:sigma2XY}
\end{align}
with $X,Y = U,J$ and $^*$ the complex conjugate.


\section{Estimation} \label{sec:MLE}
The purpose of this research is to use impedance, or equivalently voltage and current data, to estimate the potential-independent rate constants $k_{fi}$ and $k_{bi}$. 
To ensure the unique identifiability of these constants (see Section~\ref{sec:ID}), we will combine EIS data from different potentials $u_\mathrm{eq}$.
The first part of this section details the numerically stable implementation of the maximum likelihood estimator over multiple potentials.
The second part discusses other estimators that can be used to obtain an initial estimate of the rate constants.

\subsection{Maximum Likelihood Estimation}
The reaction rate constants are estimated using the sample maximum likelihood (SML) estimator \cite{Pintelon2012}.
Denoting ${\bf k}\in\mathbb{R}^{(2n+1) \times 1}$ as the vector containing the rate constants $k_{fi}$ and $k_{bi}$, the SML estimate $\hat {\bf k}_{\up{SML}}$ minimises the cost function
\begin{equation}
   V_{\up {SML}}({\bf k})
   = \sum_{u_{\mathrm{eq}} \in \mathbb{U}_{\mathrm{eq}}} \sum_{k \in \mathbb{K}_{\mathrm{exc}}} \dfrac{|\hat e(k,u_{\mathrm{eq}},{\bf k})|^2}{\hat \sigma^2_{\hat e}(k,u_{\mathrm{eq}},{\bf k})}
   \label{eq:cost}
\end{equation}
Note that this cost function is an extension of the standard SML cost, since in addition to summing over the different excited frequencies, it also sums over the different applied potentials.
Hence, the SML estimator takes measurement uncertainty into account by weighing the equation error $\hat e(k,u_{\mathrm{eq}},{\bf k})$, at each frequency bin $k \in \mathbb{K}_{\mathrm{exc}}$ and each potential $u_{\mathrm{eq}} \in \mathbb{U}_{\mathrm{eq}}$, by its variance $\hat \sigma^2_{\hat e}(k,u_{\mathrm{eq}},{\bf k})$. 

\subsubsection{Matrix implementation}

Defining $N=N_{\up{exc}} \times N_{\up{eq}}$, with $N_{\up{exc}}$ the number of excited frequencies and $N_{\up{eq}}$ the number of equilibrium potentials,
the cost function \eqref{eq:cost} can be written in matrix notation as
\begin{equation}
   V_{\up {SML}}({\bf k})
   = \hat {\bf e}^H({\bf k})\hat C_{\hat e}^{-1}({\bf k})\hat {\bf e}({\bf k})
   \label{eq:cost1}
\end{equation}
where $\hat {\bf e}({\bf k}) \in \mathbb{C}^{N\times1}$ is the vector of equation errors 
\begin{equation}
   \hat {\bf e}({\bf k}) = \mathcal A({\bf k})\hat {\bf J} - \mathcal B({\bf k})\hat {\bf U}
\end{equation}
and $\hat C_{\hat e} \in \mathbb{R}^{N\times N}$ is the covariance matrix of these equation errors
\begin{align}
   \hat C_{\hat e}({\bf k}) &= \mathcal A({\bf k})\hat C_{\hat J}\mathcal A^H({\bf k}) + \mathcal B({\bf k})\hat C_{\hat U}\mathcal B^H({\bf k}) \nonumber\\
   &- 2 \up{Re}\big(\mathcal A({\bf k})\hat C_{\hat J \hat U}({\bf k})\mathcal B^H({\bf k})\big)
\end{align}
Here, the vectors $\hat {\bf U} ,\hat {\bf J} \in \mathbb{C}^{N\times1}$ are obtained by stacking the mean voltage and current DFT spectra \eqref{eq:meanX}, 
evaluated at the excited bins $k\in\mathbb{K}_{\up{exc}}$, of the measurements at the different equilibrium potentials $u_{\mathrm{eq}} \in \mathbb{U}_{\mathrm{eq}}$.
Analogously, the diagonal matrices $\hat C_{\hat U} ,\hat C_{\hat J}\in \mathbb{R}^{N\times N}$ and $\hat C_{\hat J\hat U} \in \mathbb{C}^{N\times N}$ are constructed from the noise covariances \eqref{eq:sigma2XY}, scaled with a factor $1/M$.
Finally, the matrices $\mathcal A({\bf k}),\mathcal B({\bf k}) \in \mathbb{C}^{N\times N}$ are diagonal as well, and their diagonals are computed by respectively evaluating the denominator and the numerator of the transfer function \eqref{eq:TF} in $s=j\omega_k$, 
with $j$ the imaginary unit and $\omega_k$ for $k\in\mathbb{K}_{\up{exc}}$ the excited frequencies, for all the potentials $u_{\mathrm{eq}} \in \mathbb{U}_{\mathrm{eq}}$. \\[-1mm] 

The cost function \eqref{eq:cost1} can be rewritten as 
\begin{equation}
   V_{\up {SML}}({\bf k})
   = \hat {\boldsymbol\varepsilon}^H({\bf k})\hat {\boldsymbol\varepsilon}({\bf k})
   \label{eq:cost2}
\end{equation}
with the vector of weighted residuals
\begin{equation}
   \hat{\boldsymbol\varepsilon}({\bf k}) = \hat C_{\hat e}^{-1/2}({\bf k})\hat{\boldsymbol e}({\bf k})
\end{equation}
The cost function is minimised with the Levenberg-Marquardt algorithm, an iterative nonlinear optimization algorithm which requires the calculation of the Jacobian matrix $\hat J({\bf k})\in\mathbb{C}^{N\times (2n+1)}$,
whose columns are constructed as follows
\begin{align}
   &\hat J_{[:,j]}({\bf k}) = \frac{\partial \hat{\boldsymbol\varepsilon} ({\bf k})}{\partial { k}_j} \\
   &= \hat C_{\hat e}^{-1/2}({\bf k})\frac{\partial \hat{\boldsymbol e} ({\bf k})}{\partial { k}_j} - \frac{1}{2} \hat C_{\hat e}^{-3/2}({\bf k})\frac{\partial \hat C_{\hat e}({\bf k})}{\partial { k}_j}\hat{\boldsymbol e}({\bf k}) 
\end{align}
with
\begin{equation}
   \frac{\partial \hat {\bf e} ({\bf k})}{\partial { k}_j} 
   = \frac{\partial \mathcal A ({\bf k})}{\partial { k}_j} \hat {\bf J}
   - \frac{\partial \mathcal B ({\bf k})}{\partial { k}_j} \hat {\bf U}
\end{equation}
and
\begin{align}
   &\frac{\partial \hat C_{\hat e} ({\bf k})}{\partial { k}_j} 
   = 2\up{Re}\bigg(\frac{\partial \mathcal A ({\bf k})}{\partial { k}_j}\hat C_{\hat J}\mathcal A^H({\bf k}) \nonumber\\
   & + \frac{\partial \mathcal B ({\bf k})}{\partial { k}_j}\hat C_{\hat U}\mathcal B^H({\bf k}) 
   - \frac{\partial \mathcal A ({\bf k})}{\partial { k}_j}\hat C_{\hat J\hat U}\mathcal B^H({\bf k})\nonumber \\
   & - \mathcal A({\bf k})\hat C_{\hat J\hat U}\frac{\partial \mathcal B^H ({\bf k})}{\partial { k}_j}\bigg)
\end{align}
for $j = 1,\dots,2n+1$.
The diagonal matrices ${\partial \mathcal A ({\bf k})}/{\partial { k}_j},{\partial \mathcal B ({\bf k})}/{\partial { k}_j}\in \mathbb{C}^{N\times N}$
are caculated by taking the partial derivatives of the denominator and the numerator of the transfer function \eqref{eq:TF} to the rate constants $k_j$.
These partial derivatives are again evaluated in $s=j\omega_k$ for $k\in\mathbb{K}_{\up{exc}}$, and stacked for all $u_{\mathrm{eq}} \in \mathbb{U}_{\mathrm{eq}}$.

\subsubsection{Numerically stable implementation}

To ensure that the estimated rate constants are real, 
the real and imaginary parts of the residual vector and the Jacobian matrix are stacked on top of each other,
\begin{equation}
   {X}_{\up{re}}({\bf k}) 
   = \begin{bmatrix} \up{Re}(X({\bf k})) \\ \up{Im}(X({\bf k})) \end{bmatrix}
   \text{ for } X = \hat{\boldsymbol \varepsilon},\hat J
\end{equation}
so that $\hat{\boldsymbol \varepsilon}_{\up{re}}({\bf k})\in \mathbb{R}^{2N\times1}$ and $\hat J_{\up{re}}({\bf k})\in \mathbb{R}^{2N\times(2n+1)}$.
Furthermore, to increase numerical stability, the columns of the extended Jacobian are scaled with their 2-norms, 
\begin{equation}
   \hat J_{\up{S}}({\bf k}) =\hat  J_{\up{re}}({\bf k})S^{-1}({\bf k})
\end{equation}
where $S({\bf k})\in\mathbb{R}^{(2n+1)\times(2n+1)}$ is a diagonal matrix containing these 2-norms.
At each iteration, the Levenberg-Marquardt algorithm with tuning parameter $\lambda$  then determines the scaled update $\Delta{\bf k}_{\up{S}}$ by solving 
\begin{equation}
   \left(\!\hat J_{\up{S}}^\top({\bf k})\hat J_{\up{S}}({\bf k}) + \lambda I_{2n+1}\!\right)\Delta {\bf k}_{\up{S}} = -\hat J_{\up{S}}^\top({\bf k})\hat{\boldsymbol \varepsilon}_{\up{re}}({\bf k})
\end{equation}
with singular value decomposition (SVD) of the scaled Jacobian.
The actual update of the estimated rate constants is then found as 
\begin{equation}
   \Delta {\bf k} = S^{-1}({\bf k})\Delta {\bf k}_{\up{S}}
\end{equation}
Next to scaling the columns of the Jacobian, numerical stability is also improved by scaling the frequency axis.
Here, the median frequency $\omega_m$ of the logarithmic excited frequency band will be used as a scaling factor.
The denominator and the numerator of the transfer function \eqref{eq:TF} will then become
\begin{align}
   \tilde A(s) &= \sum_{i=0}^n \tilde a_i\left(\frac{s}{\omega_m}\right)^i \text{ with } \tilde a_n = 1\\
   \tilde B(s) &= \sum_{i=0}^n \tilde b_i\left(\frac{s}{\omega_m}\right)^i
\end{align}
If the goal would have been to estimate the coefficients of the transfer function, this means that instead of $a_i$ and $b_i$ themselves, 
one would actually estimate the scaled coefficients $\tilde a_i = a_i\omega_m^{i-n}$ and $\tilde b_i = b_i\omega_m^{i-n}$ \cite{Pintelon2012}. 
However, we are not estimating these transfer function coefficients, but the reaction rate constants.
Fortunately, there is a specific structure in the relationship between the coefficients and the rate constants.
Namely, the coefficients are homogenous polynomials in the rate constants, i.e. the expression of each coefficient consists exclusively of terms in the rate constants of exactly the same degree
\begin{align}
   a_i &\sim O(k^{n-i}) \label{eq:ai_unscaled}\\
   b_i &\sim O(k^{n-i+1}) \label{eq:bi_unscaled}
\end{align}
Therefore, we propose to estimate the scaled rate constants $\tilde k_j = k_j/\omega_m$. 
Writing down the original relationship between the coefficients and the rate constants with their scaled values gives 
\begin{align}
  \tilde a_i &= a_i\omega_m^{i-n} \sim O(\tilde k^{n-i}) = O( k^{n-i})\omega_m^{i-n}  \label{eq:ai_scaled}\\
  \tilde b_i &= b_i\omega_m^{i-n} \sim O(\tilde k^{n-i+1}) = O( k^{n-i+1})\omega_m^{i-n-1} \label{eq:bi_scaled}
\end{align}
From comparing \eqref{eq:ai_unscaled}-\eqref{eq:bi_unscaled} with \eqref{eq:ai_scaled}-\eqref{eq:bi_scaled}, it follows that 
estimating the rate constants $k_j$ with an unscaled frequency axis, is equivalent with estimating the scaled rate constants $\tilde k_j$ with a scaled frequency axis,
if the resulting numerator $\tilde B(s)$ is multiplied with a factor $\omega_m$.\\[-1mm]

A similar derivation can be made for the partial derivatives of the denominator and the numerator to the rate constants $k_j$.
Here, we get that
\begin{align}
   \frac{\partial a_i}{\partial k_j} &\sim O(k^{n-i-1}) \label{eq:dai_unscaled}\\
   \frac{\partial b_i}{\partial k_j} &\sim O(k^{n-i}) \label{eq:dbi_unscaled}
\end{align}
such that
\begin{align}
   &\!\!\frac{\partial \tilde a_i}{\partial \tilde k_j} = \frac{\partial a_i}{\partial k_j}\omega_m^{i-n+1} 
   \sim O(\tilde k^{n-i-1}) = O( k^{n-i-1})\omega_m^{i-n+1}  \label{eq:dai_scaled}\\
   &\!\!\frac{\partial \tilde b_i}{\partial \tilde k_j} = \frac{\partial b_i}{\partial k_j}\omega_m^{i-n+1} 
   \sim O(\tilde k^{n-i}) = O( k^{n-i})\omega_m^{i-n}  \label{eq:dbi_scaled}
\end{align}
Hence, also the partial derivatives of the numerator have to be multiplied by $\omega_m$. 

\subsubsection{Properties of the SML estimation}

The SML estimate $\hat {\bf k}_{\up{SML}}$ is consistent if the number of measured periods $M \geq 4$ \cite{Pintelon2012}.
This means that the estimate converges to the true rate constants ${\bf k}_0$ when the number of excited frequencies and equilibrium potentials increases. 
However, the SML estimate is not asymptotically efficient.
Indeed, for $M\geq7$ \cite{Pintelon2012} the asymptotic covariance matrix can be approximated by 
\begin{equation}
   \up{Cov}(\hat {\bf k}_{\up{SML}}) \!\approx\! \frac{M\!-\!1}{M\!-\!3} \left[2\up{Re} (\hat J^H(\hat {\bf k}_{\up{SML}})\hat J(\hat {\bf k}_{\up{SML}}))\right]^{-1}
   \label{eq:acov}
\end{equation}
which does not reach the Cramér-Rao lower bound, given by the inverse Fisher information matrix
\begin{equation}
   \up{Fi}^{-1}({\bf k}_0) = \left[2\up{Re}(J_0^H({\bf k}_0)J_0({\bf k}_0))\right]^{-1}
   \label{eq:CR}
\end{equation}
Here, the Jacobian $J_0({\bf k}_0)$ is constructed from the noiseless data instead of the sample means and sample noise covariances, and evaluated in the true rate constants.

\subsection{Initial Value Estimation}
Not only is the impedance model highly nonlinear in the rate constants, but the orders of magnitude of these rate constants are also very diverse.
Finding good enough starting values, to avoid the nonlinear optimization ending up in a local minimum, is thus extremely important. 
Therefore, this paper first performs a global search of the parameter space to generate good initial values, 
followed by a local search with the Levenberg-Marquardt algorithm to converge to the minimum fast and accurately. \\[-1mm]

Three nature inspired global optimization algorithms are considered: 
evolution strategies (ES) and genetic algorithm (GA), both based on natural evolution, and particle swarm optimization (PSO), based on the movement of animals in a swarm. 
Note that since these methods inherently use random variables, different runs will lead to different results.
This stochastic character is exactly what allows them to escape from local minima by forgetting the corresponding parameter values, at the cost of a reduced convergence speed \cite{nonlinear2020}.\\[-1mm]

The algorithms are capable of examining a vast search space by working with a whole population of individuals, which are the possible solutions.
The population evolves over the iterations by applying some method-specific operators.
The final result is the best individual, i.e. the solution that corresponds to the smallest value of the cost function, over all the iterations.
A short overview of the three techniques is given below. 

\subsubsection{Evolution Strategies (ES)}
\label{sec:ES}
In evolution strategies, an individual $i$ contains not only the parameter values ${\boldsymbol x}_i$, but also the \textit{strategy parameters} ${\boldsymbol\sigma}_i$, which are the standard deviations of the parameter step sizes.
One of the main strengths of the ES algorithm lies in its \textit{self-adaptation}: the strategy parameters, which influence the optimization algorithm itself, are also subject to optimization. 
Each iteration, a population of $\mu$ parents produces $\lambda$ children by applying the genetic operators recombination and mutation, of which $\mu$ children are then selected to be the next generation of parents.
We use a discrete recombination of the parameters, meaning that each child inherits some parameters from one parent and some from another,
and an intermediate recombination of the strategy parameters, meaning that the standard deviations are simply averaged between parents. 
After that, the strategy parameters are mutated according to a log-normal distribution, and the parameters are mutated according to a Gaussian distribution as follows
\begin{align}
   {\boldsymbol\sigma}_i =& {\boldsymbol\sigma}_i e^{\Delta\boldsymbol\sigma_i + \Delta\boldsymbol\sigma_\up{pop}} \text{ with } \Delta{\boldsymbol\sigma}_i \sim \mathcal N({\bf{0}},\tau_iI_n) \nonumber \\
   & \text{ and } \Delta{\boldsymbol\sigma}_\up{pop} \sim \mathcal N({\bf{0}},\tau_\up{pop}I_n)\\
   {\boldsymbol x}_i =&{\boldsymbol x}_i + \Delta {\boldsymbol x}_i \text{ with } \Delta{\boldsymbol x}_i \sim \mathcal N({\bf{0}},\up{diag}({\boldsymbol\sigma}_i))
\end{align}
Here, ${\bf{0}}\in\mathbb{R}^n$ and  $I_n\in\mathbb{R}^{n\times n}$ respectively denote the zero vector and the identity matrix, with $n$ the number of parameters.
The learning rates of the individual and the population are set to \cite{ESoriginal}
\begin{equation}
   \tau_i = \frac{1}{\sqrt{2\sqrt{n}}} \quad \tau_\up{pop} = \frac{1}{\sqrt{2n}}
\end{equation}
Next, the cost function is evaluated in the feasible individuals, i.e. those that satisfy the user-defined upper and lower bounds ${\boldsymbol x}_\up{min} \leq {\boldsymbol x}_i \leq {\boldsymbol x}_\up{max}$.
To ensure that feasible solutions are always selected before infeasible ones, and that infeasible solutions are ranked based on the extent to which they violate the constraints,  
we use the adapted cost function proposed by \cite{Constraints2000}
\begin{align}
   &\tilde V_\up{SML}({\boldsymbol x}_i) \nonumber\\ &= 
   \begin{cases} 
   V_\up{SML}({\boldsymbol x}_i) & \text{if } {\boldsymbol x}_i  \text{ feasible} \\ 
   \begin{aligned} &V_\up{worst} + {\bf 1} \cdot \max ({\bf 0},{\boldsymbol x}_{i}-{\boldsymbol x}_{\max}) \\
   & + {\bf 1} \cdot \max ({\bf 0},{\boldsymbol x}_{\min}-{\boldsymbol x}_{i}) \end{aligned} & \text{otherwise}  
   \end{cases}
\end{align}
where $V_\up{worst}$ is the cost function value of the worst feasible individual in the population, and $\cdot$ denotes the inproduct.
Finally, the $\mu$ best individuals, i.e. those with the lowest cost, are selected.
We opted for a $(\mu,\lambda)$-selection, which entails that the selection is performed only on the children, as opposed to $(\mu+\lambda)$-selection, where parents compete with their children. 
The former allows for a more global search, since successful parents can be forgotten, so that it is possible to escape from local minima.  
As a ratio $\mu/\lambda \approx 1/7$ is recommended in literature \cite{nonlinear2020},\cite{ESoriginal}, this work uses a $(15,100)$-selection scheme.

\subsubsection{Genetic Algorithm (GA)}
The genetic algorithm is another evolutionary algorithm, but now an individual only contains the parameter values themselves.
In the original GA, these parameters are coded as a bitstring, similar to the way genetic information in DNA is coded with only four symbols A, T, C and G.
Since the rate constants are real numbers, however, it would require a significant amount of bits to represent them accurately, making the size of the individuals very large.
Hence, this work uses the real-coded GA implementation of MATLAB\textsuperscript{\circledR}. \\[-1mm]

The algorithm works as follows. 
Each iteration, a number of parents are selected with repetition out of a population of $\lambda$ individuals.
The selected parents will then create the next generation of $\lambda$ individuals by recombination and mutation. 
Contrary to ES, selection is probabilistic, meaning that the best individuals are more likely, but not guaranteed, to be selected for procreation.
We use a roulette wheel selection, where random numbers drawn from a uniform distribution between 0 and 1 select sections of a wheel, representing different individuals, as parents.
The area of each wheel section equals the probability that the corresponding individual is selected.
To make the probabilities of being chosen independent of the exact cost function definition, we choose them inversely proportional to the square root of the rank $r$ of the individuals, where $r=1$ corresponds to the individual with the lowest cost and $r=\lambda$ to the one with the highest cost. \\[-1mm] 

Unlike in ES, where genetic diversity is primarily introduced by mutation, the GA's main genetic operator is recombination, called crossover.
The probability that crossover will occur between two selected parents, the crossover rate $P_c$, is thus typically rather large, e.g. 0.6--1 \cite{nonlinear2020},\cite{ESoriginal}, 
whereas the probability that a real-valued individual will be mutated, the mutation rate $P_m$, is quite small, e.g. 0.01--0.4 \cite{GAbook}.   
Note that while usually the selected parents are first recombined with probability $P_c$, after which the resulting children are mutated with probability $P_m$, 
MATLAB makes these two operations mutually exclusive, such that children are either created by recombination or by mutation of the parents.
Consequently, MATLAB only allows to set the crossover rate, and the mutation rate is automatically computed as $P_m = 1-P_c$.
In the original GA, crossover corresponds to discrete recombination in ES, i.e. the parameters of two parents are interchanged to form a child. 
The problem with this method is that no new information is introduced, as the randomly initialised parameters are passed on from generation to generation \cite{GAbook}. 
Therefore, we chose a randomly weighted intermediate recombination of the parameters.
The remaining parents are mutated by adding steps in random directions. 
The adaptive step sizes are such that the resulting children satisfy the bounds.
The population size is 100, of which the 5 best individuals, called elites, are immediately passed on to the next generation without undergoing crossover or mutation \cite{GAbook}. \\[-1mm]

\begin{figure}[H]
\centering
\includegraphics[scale = 0.75]{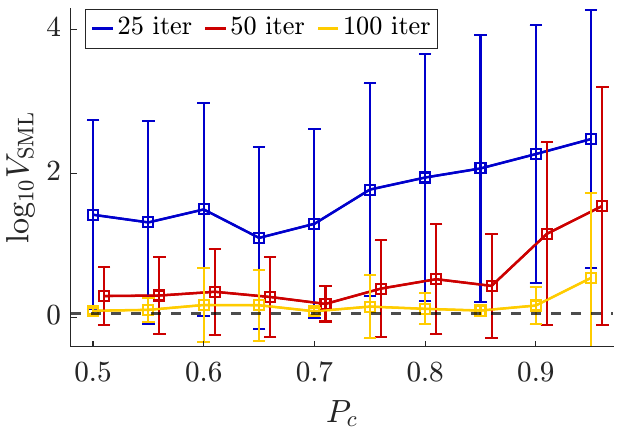} 
\caption{Means and standard deviations over 40 realisations of the logarithmic cost in the estimated rate constants $\log_{10} V_\up{SML}(\hat {\bf k})$, after 25 (blue), 50 (red) and 100 (yellow) iterations of the genetic algorithm, for varying crossover rates $P_c$ between 0.6 and 1.
The black dotted line indicates the cost in the true rate constants $\log_{10} V_\up{SML}({\bf k}_0)$.
Simulation of the second order impedance model with 40~dB noise, $N_\mathrm{exc} = 90$ excited 
\vspace{-4mm}}
\label{fig:XoverFraction}
\end{figure}

To find a good value of $P_c$, and thus also of $P_m$, MATLAB's GA is run 40 times for different crossover rates between 0.6 and 1. 
The means and the standard deviations of the cost function after 25, 50 and 100 iterations are plotted in \textbf{Figure}~\ref{fig:XoverFraction}. 
A crossover rate $P_c = 0.7$ yields the best result, as the cost converges fast (after 50 iterations) and accurately (with a small standard deviation) to the cost in the true rate constants.
Remark that choosing a smaller $P_c$ means relying more on mutation to introduce genetic diversity, such that the real-coded GA starts behaving as the evolution strategies algorithm.

\subsubsection{Particle Swarm Optimization (PSO)}
In particle swarm optimization, individuals are called particles and the population is called the swarm.
An individual $i$ is characterised by its position ${\boldsymbol{x}}_i$, i.e. the parameter values, and its velocity ${\boldsymbol{v}}_i$, i.e. the parameter step sizes.
Each iteration, the velocity is adapted based on the historical best positions of the individual itself and of the whole population. 
Then the position is updated by adding the velocity \cite{PSOoriginal} 
\begin{align}
   \!\!\! {\boldsymbol{v}}_i &= w{\boldsymbol{v}}_i + c_1{\bf{r}}_1 \!\circ\! ({\boldsymbol{x}}_{i}^{\up{best}}\!\!-\!{\bf{x}}_i) + c_2{\bf{r}}_2  \!\circ\! ({\boldsymbol{x}}_{\up{pop}}^{\up{best}}\!-\!{\bf{x}}_i) \!\!\!\!\label{eq:velocity}\\
   \!\!\! {\boldsymbol{x}}_i &= {\boldsymbol{x}}_i+{\boldsymbol{v}}_i
\end{align}
Here, $w$ is referred to as the inertia weight, and $c_1$ and $c_2$ are the learning rates of the individual and the population respectively. 
The vectors ${\bf{r}}_1$ and ${\bf{r}}_2$ contain random numbers drawn from a uniform distribution between 0 and 1, and $\circ$ denotes the elementwise product. 
This paper uses the MATLAB\textsuperscript{\circledR} implementation of PSO, in which the inertia weight $w$ is automatically adjusted over the iterations. 
Initially, setting $w>1$ allows to explore the search space. 
Afterwards, gradually decreasing the value to $w<1$ leads to convergence. 
A common choice in literature for the learning rates is $c_1 = c_2 = 1.49$ \cite{PSOconstriction_vs_inertia},\cite{PSOexample}, which guarantees fast convergence once the inertia weight becomes small enough ($w<0.75$). 
Like in ES and GA, the population size is set to 100. \\[-1mm] 



The original PSO algorithm uses a global topology in which all particles can communicate with all others, such that they all share the same global best position ${\boldsymbol{x}}_{\up{pop}}^{\up{best}}$. 
As this can cause premature convergence to a local minimum, an alternative is to use a local topology in which each particle can only communicate with a subset of particles, called the neighbourhood.
That way, the global best solution ${\boldsymbol{x}}_{\up{pop}}^{\up{best}}$ in \eqref{eq:velocity} is replaced by the local best solution ${\boldsymbol{x}}_{\up{neighbour},i}^{\up{best}}$ found in the neighbourhood of the individual $i$.
For a more global search, this work starts with only two neighbouring particles. 
MATLAB's PSO then gradually increases the size of the neighbourhood over the iterations until it encompasses the entire swarm.


\subsubsection{Comparison of the global optimization algorithms}
The global optimization algorithms are used to estimate the rate constants of a hematite photoanode (see Section~\ref{sec:Simulation}), from simulated data of the second order impedance model.
For $N_\mathrm{eq} = 4$ potentials between 1.4~V and 1.46~V, $N_\mathrm{exc} = 90$ frequencies are excited within the frequency band $[10^{-2},10^2]$~Hz. 
The resulting voltage and current signals are disturbed by 40~dB noise.\\[-1mm]

For the sake of comparison, all the algorithms were run with a population size of 100, 
until either the cost became less than the minimum value $V_{\up{min}} = V_\up{SML}({\bf k}_0)$, or the number of iterations exceeded 500.
For each algorithm, we performed 100 Monte-Carlo runs with different starting values, randomly generated within the user-defined bounds $[10^{-30},10^{-15}]$. 
Due to the large numerical range of the rate constants (see Table~\ref{table:RateConstants}), the global optimization methods actually search for the logarithmic values of the rate constants $\log_{10}{\bf k}$. 
Therefore, the initial values $\log_{10} {\bf k}_\mathrm{init}$ are drawn from a uniform distribution between $-30$ and $-15$.\\[-1mm]

Table~\ref{table:GlobalOptimization} shows the number of runs that converged to the true rate constants, as well as the minimum, maximum, median, mean and standard deviation of the number of cost function evaluations over the converged runs.
While GA often gets stuck in local minima, both ES and PSO have a high convergence rate. 
However, evolution strategies needs less function evaluations, and is thus faster.  
This is due to both the efficient constraint handling, i.e. the cost in not explicitly calculated if the constraints are violated, and the fact that ES requires fewer iterations in general.\\[-1mm]

\begin{figure}[H]
\centering
\includegraphics[scale = 0.75]{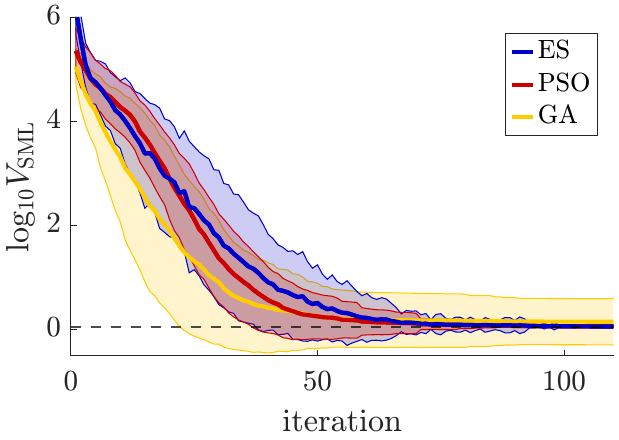} 
\caption{Means and standard deviations over 100 realisations of the logarithmic cost in the estimated rate constants $\log_{10} V_\up{SML}(\hat {\bf k})$, over the iterations of the different global optimization algorithms: ES (blue), PSO (red) and GA (yellow).
The black dotted line indicates the cost in the true rate constants $\log_{10} V_\up{SML}({\bf k}_0)$.
\vspace{-4mm}}
\label{fig:CostVsIteration}
\end{figure}

\textbf{Figure}~\ref{fig:CostVsIteration} shows the means and the standard deviations of the logarithmic cost in the estimated rate constants $\log_{10} V_\up{SML}(\hat {\bf k})$ over the iterations. 
It is clear that the GA converges faster, but since it often ends up in a local minimum, the converged average cost does not equal the minimum cost, and the uncertainty remains quite large.
Also note that the PSO cost is guaranteed to decrease over the iterations, as it is calculated based on the historical best position of the population.
The ES cost, on the other hand, sometimes increases as well, since all the individuals are continuously mutated, and this does not necessarily bring them closer to the true rate constants.
However, this does allow the ES to escape from local minima.


\begin{table}[H] 
\centering
\caption{Comparison of global optimization algorithms. Number of converged runs out of 100 Monte-Carlo runs, and the statistics of the number of cost function evaluations in these converged runs.}
\label{table:GlobalOptimization}
\begin{tabular}{c|c|c|c|c} 
 \multicolumn{2}{c|}{Algorithm} & ES & PSO & GA  \\
 \hline
\multicolumn{2}{c|}{\# converged runs} &     97  &  96  &  64\\
 \hline
\multirow{5}{*}{\begin{turn}{90} \shortstack{\# function \\ evaluations}  \end{turn} }  
& min  &       5437  &   7200   &  5425\\
&max   &      11455 &   40900  &  44755\\
&median   &      7851  &   12300  &  17965\\
&mean  &      8012   &  13323  &  19856\\
&std   &      1192  &   4716   &  8490
\end{tabular}
\end{table}

\section{Identifiability}
\label{sec:ID}

This section investigates under which conditions the rate constants can be estimated uniquely from the available impedance data.
Since this impedance data depends on the excited frequency band and on the applied potential, 
the influence of both these factors on the identifiability of the rate constants will be examined. 
\\[-1mm]




First of all, remark that the $n$th order impedance model contains only $2n+1$ reaction rate constants, 
namely the forward rate constants $k_{fi}$ for $i = 1,\dots,n+1$ and the backward rate constants $k_{bi}$ for $i = 1,\dots,n$. 
Moreover, the $n$th order transfer function \eqref{eq:TF} also has $2n+1$ coefficients,   
i.e. $a_i$ for $i=0,\dots,n-1$ and $b_i$ for $i=0,\dots,n$.
Using relation \eqref{eq:admittance}, one can write down the analytical expressions of these transfer function coefficients as a function of the rate constants.
The result is a nonlinear system of $2n+1$ equations in the $2n+1$ rate constants.
Hence, if it is possible to identify the coefficients of the transfer function, or, equivalently, its gain, poles and zeros, the rate constants are (locally) identifiable as well. 

\subsection{Identifiability in a given frequency band}

Especially at low potentials, the hole transfer reactions rarely occur. 
A part of the dynamics is therefore found at low, practically unmeasurable frequencies.
The desorption reaction, on the other hand, is a fast reaction, resulting in high-frequency dynamics.
As a consequence, actual measurements, where the frequency band is restricted, will most likely not capture the full behaviour of the fourth order or even the reduced order models.
Hence, when tackling the question of identifiability, the frequency band in which the impedance data is available has to be taken into account. \\[-1mm]

In an EIS measurement, the minimum frequency is determined by the period $T$ of the multisine excitation signal, 
$f_\up{min} = 1/T$ (see Section~\ref{sec:FreqDomainData}).
The maximum frequency is restricted by the sampling frequency $f_{\up{s}}$ of the measurement device, since it has to satisfy the Nyquist theorem $f_{\up{max}}<f_{\up{s}}/2$. \\[-1mm]

\begin{figure}[H]
\centering
\includegraphics[scale = 0.75]{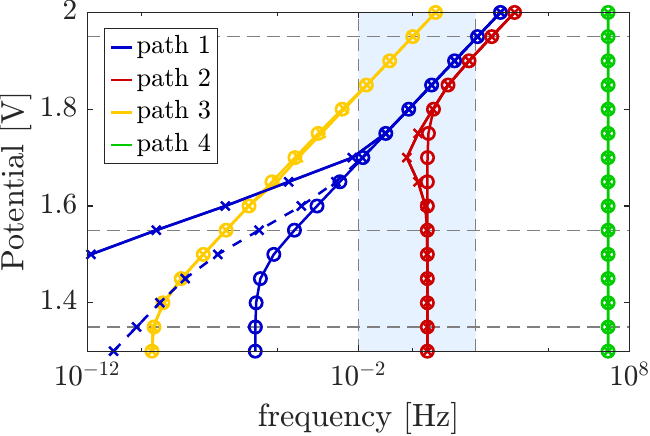} 
\caption{Poles (crosses) and zeros (circles) of the fourth order impedance model as a function of potential. 
Reduction of the model order eliminates the higher order pole-zero paths, marked with different colours.
For example, the model of order $n=1$ only contains the blue path, the model of order $n = 2$ contains the blue and the red path, etc.
Except for the poles of the first order model (dashed blue line), the poles and zeros of the reduced order models approximately coincide with those of the fourth model.
The horizontal dashed lines indicate the potential boundaries of the different model orders (see Figure~\ref{fig:SpeciesCoverage}).
The blue region corresponds to the measurable frequency band [10~mHz, 200~Hz] considered in Figure~\ref{fig:BodePlotsReducedOrder}.
\vspace{-4mm}}
\label{fig:PathsPolesZeros}
\end{figure}

Consider \textbf{Figure}~\ref{fig:PathsPolesZeros}, 
which illustrates how the frequencies of the poles and zeros of the fourth order impedance model vary with the applied potential.
Note here that the poles and zeros of the impedance are respectively the zeros and poles of the admittance \eqref{eq:TF}.
Reducing the model order to $n<4$ eliminates the higher order pole-zero paths. 
In other words, the $n$th order impedance model consists of paths 1 up to $n$.
For example, the first order model only contains the blue path, the second order model contains the blue and the red path, and so on.
In practice, the paths of the reduced order models differ slightly from those of the fourth order model, but this difference is generally negligible: 
The zeros of the reduced order models coincide with those of the fourth order model (relative error less than 2\%). 
The poles correspond reasonably well (relative error less than 25\%), with the exception of those of the first order model, which are indicated separately with the dashed blue line. \\[-1mm]

Figure~\ref{fig:PathsPolesZeros} clearly shows that at low potentials, the blue, first order poles and zeros are found at extremely low frequencies of less than 1~$\mu$Hz.
Therefore, the first order model will only be used at very low frequencies (and low potentials), to model the rare first formations of OH$^*$.
Also, the fourth order poles and zeros, in green, coincide and thus cancel each other out at 16~MHz, which makes them unidentifiable.
As a result, except at very high frequencies (and high potentials), 
the fourth order model will always be approximated by the third order model.
Physically, this can be explained by the fact that the oxygen desorption rate $K_{f5}$ is very large compared to the other reaction rates.
Hence, once $\up{O}_2^*$ is formed, it immediately disappears again and thus its fractional coverage remains negligible (see Figure~\ref{fig:SpeciesCoverage}).
Consequently, it is virtually impossible to identify the rate constants $k_{b4}$ and $K_{f5}$, since they only appear in the fourth order model. \\[-1mm]

Figure~\ref{fig:PathsPolesZeros} also clarifies why at least a model order of two is required within the frequency band of 10~mHz to 200~Hz (see Figure~\ref{fig:BodePlotsReducedOrder}). 
This is because the red, second order pole-zero path is the dominant one in this band.
In general, a lot of the pole-zero pairs coincide almost perfectly, thus quasi-cancelling each other out. This limits the dynamic range of the impedance (i.e. the difference between the largest and the smallest impedance value) and significantly complicates the identification. 
Only between 1.55~V and 1.95~V, where the third order model is used, do the pole and zero paths begin to diverge from each other, thus giving rise to more complex dynamics. 
This potential range is therefore the most informative to do measurements. \\[-1mm]


Consider again the Bode plots in Figure~\ref{fig:BodePlotsReducedOrder}, which depict the impedance spectra between 10~mHz and 200~Hz.
Remembering that poles and zeros respectively result in a decrease and an increase of both magnitude and phase, it is clear that at 1.5~V and 1.6~V, 
only one pole and one zero are found inside this frequency band. 
From Figure~\ref{fig:PathsPolesZeros}, it follows that these are the second order pole and zero. 
Hence, taking into account the gain, there are only three transfer function elements available to identify respectively five rate constants for the second order model at 1.5~V, and seven rate constants for the third order model at 1.6~V.
At 1.7~V, the first order zero enters the band, so there are four elements to identify seven rate constants, which is still not enough.
Finally, at 1.8~V, we see both the first order pole and zero, which almost coincide and thus cancel each other out, and the second order pole and zero, inside the band. 
Also here, it will not be possible to identify the seven rate constants from these five elements.
Fortunately, this issue will be solved by combining impedance measurements at multiple potentials.

\subsection{Identifiability over multiple potentials}

Assuming a wide enough frequency band, we already established that the rate constants are locally identifiable, as finding the rate constants is equivalent to solving a nonlinear system of $2n+1$ equations in $2n+1$ unknowns. 
Mathematically, such a nonlinear system has a myriad of solutions, of which only a few are physically feasible.
But what if data from $N_\mathrm{eq}=2$ potentials is combined? In that case, there $2\times(2n+1)$ transfer function coefficients to identify the same $2n+1$ rate constants, and only one solution will remain.
In other words, in a wide enough frequency band, all rate constants are globally identifiable if at least two potentials are combined. 
However, more potentials help in reducing the uncertainty, as is illustrated in Section~\ref{sec:Simulation}.

\section{Simulation} \label{sec:Simulation}
The OER model is simulated for a hematite (Fe$_2$O$_3$) photoanode. 
The model parameters for the hematite-water interface are listed in Table~\ref{table:ModelParameters}.
The rate constants to be estimated are given in Table ~\ref{table:RateConstants}.\\[-1mm]

EIS data is simulated in the frequency domain, by applying multisine potential excitations and calculating the current density responses from the impedance model.
The multisines have an RMS value of 1~mV, and consist of $N_\mathrm{exc} = 90$ excited frequencies that are logarithmically distributed between 10~mHz and 100~Hz.
Data from $N_\mathrm{eq} = 4$ equilibrium potentials is combined, in steps of 0.02~V between 1.4~V and 1.46~V for the second order model, and between 1.7~V and 1.76~V for the third order model.
The voltage and current signals are simulated for $M = 10$ periods of $T = 100$~s, 
at a sampling frequency $f_\mathrm{s} = 250$~Hz, 
and are disturbed by zero-mean Gaussian noise. 
The simulations are performed with a signal-to-noise ratio (SNR) of 40~dB. 
The SNR is defined as
\begin{equation}
   \mathrm{SNR}_{\mathrm{dB}} = 20 \log_{10}\left(\dfrac{\mathrm{RMS}(x)}{\sigma_x}\right)
\end{equation}
where the signal $x = u,j$ and $\sigma_x$ denotes the standard deviation of the noise. \\[-1mm]

\begin{table*}[htb!]
\centering
\caption{Model parameters for hematite-water interface, based on \cite{george2019}.}
\label{table:ModelParameters}
\begin{tabular}{c l c c} 
 \hline
 $e$     & Elementary charge  & $1.60\!\cdot\!10^{-19}$   & C\\
 $k_B$   & Boltzmann constant & $1.38\!\cdot\!10^{-23}$   & J/K\\
 $T$     & Temperature        & $298$                 & K\\
 \hline
 pH & pH of the electrolyte & $13.8$ & - \\ 
 $x_{\up{OH}}$   & Mole fraction hydroxide ions & $0.011$ &-\\
 $x_{\up{H_2O}}$ & Mole fraction water & $0.989$ &-\\
 \hline
 $p_{s0}$ & Density of holes on the surface in the dark 
 & $3.05\!\cdot\!10^{-12}$ &cm$^{-3}$\\
 $N_V$ & Density of states in the valence band 
 & $1\!\cdot\!10^{22}$ &cm$^{-3}$\\
 $N_0$ & Density of adsorption sites on the surface 
 & $1.22\!\cdot\!10^{16}$ &cm$^{-3}$\\
 \hline
 $K_{f5}$ & Oxygen desorption rate & $1\!\cdot\!10^{8}$ & 1/s 
\end{tabular}
\end{table*}

\begin{table}[H]
\centering
\caption{Forward and backward rate constants in cm$^4$/s for hematite-water interface, based on \cite{george2019}.}
\label{table:RateConstants}
\begin{tabular}{c|c|c|c|c} 
 $i$ & 1 & 2 & 3 & 4 \\
 \hline&&&\\[-3.5mm]
 $\!\!k_{fi}\!\!$ & $\!\!4.69\!\cdot\!10^{-17}\!\!\!$  & $\!\!4.40\!\cdot\!10^{-20}\!\!\!$ & $\!\!1.56\!\cdot\!10^{-16}\!\!\!$ & $\!\!1.47\!\cdot\!10^{-19}\!\!\!$\\
 $\!\!k_{bi}\!\!$ & $\!\!9.80\!\cdot\!10^{-28}\!\!\!$ & $\!\!2.19\!\cdot\!10^{-21}\!\!\!$ & $\!\!1.48\!\cdot\!10^{-31}\!\!\!\!$ & $\!\!6.70\!\cdot\!10^{-51}\!\!\!$
\end{tabular}
\end{table}

\subsection{Estimation results}
The complete estimation procedure was performed for 40 Monte-Carlo realisations of the random noise.
Each realisation, the initial values of the rate constants $\hat{\bf k}_\mathrm{init}$ are estimated with the evolution strategies algorithm $(\mu = 15,\lambda = 100$, see Section~\ref{sec:ES}), starting from different random initial values within the bounds ($[10^{-30},10^{-15}]$ for the second order model, and $[10^{-35},10^{-15}]$ for the third order model).
Then, the sample maximum likelihood estimate $\hat{\bf k}_\mathrm{SML}$ is obtained with the Levenberg-Marquardt algorithm. \\[-1mm]

For the second order model, one of the 40 runs is discarded since the initial estimation got stuck in a local minimum. 
For the third order model, only 25 runs converged within the maximum number of iterations.
This lower convergence rate can be attributed to the fact that the considered frequency band does not fully capture the dynamics of the third order pole and zero (see Figure~\ref{fig:PathsPolesZeros}). \\[-1mm]

Table~\ref{table:estimation2} and Table~\ref{table:estimation3} respectively show the estimated rate constants of the second and third order model, averaged over the converged runs, as well as the corresponding relative errors
\begin{equation}
   {\varepsilon}_j = \frac{|\hat{k}_j-{k}_{0j}|}{|{k}_{0j}|}
\end{equation}
The initial estimates $\hat {\bf k}_{\up{init}}$ are already quite close to the true values.
As expected, however, the maximum likelihood estimates $\hat {\bf k}_{\up{SML}}$ are even better (smaller relative error), except for the negative $k_{f3}$ in the second order model, which is physically impossible.
Also note that the initial estimate of $k_{f3}$ falls outside the bounds $[10^{-30},10^{-15}]$.
This is because these bounds are actually imposed on the scaled rate constants $\tilde k_j = k_j/\omega_m$. \\[-1mm]

\begin{table}[H]
\centering
\caption{Initial estimate $\hat {\bf k}_{\up{init}}$ and SML estimate $\hat {\bf k}_{\up{SML}}$, as well as the corresponding relative errors, averaged over the 39 converged Monte-Carlo runs of the second order model.}
\label{table:estimation2}
\begin{tabular}{c|c|c|c|c} 
& $\hat {\bf k}_{\up{init}}$ & ${\boldsymbol \varepsilon}_{\up{init}}$ & $\hat {\bf k}_{\up{SML}}$ & ${\boldsymbol\varepsilon}_{\up{SML}}$ \\ \hline &&&&\\[-3.5mm]
$k_{f1}$ & $4.69\!\cdot\!10^{-17}$ & $2.04\!\cdot\!10^{-5}$ & $4.69\!\cdot\!10^{-17}$ & $5.90\!\cdot\!10^{-6}$ \\
$k_{b1}$ & $9.80\!\cdot\!10^{-28}$ & $7.54\!\cdot\!10^{-5}$ & $9.80\!\cdot\!10^{-28}$ & $1.89\!\cdot\!10^{-5}$ \\
$k_{f2}$ & $4.32\!\cdot\!10^{-20}$ & $1.89\!\cdot\!10^{-2}$ & $4.41\!\cdot\!10^{-20}$ & $5.51\!\cdot\!10^{-4}$ \\
$k_{b2}$ & $2.27\!\cdot\!10^{-21}$ & $3.42\!\cdot\!10^{-2}$ & $2.21\!\cdot\!10^{-21}$ & $8.02\!\cdot\!10^{-3}$ \\
$k_{f3}$ & $1.21\!\cdot\!10^{-14}$ & $7.69\!\cdot\!10^{1}$ & $-5.12\!\cdot\!10^{-13}$ & $3.29\!\cdot\!10^{3}$
\end{tabular}
\end{table}

\begin{table}[H]
\centering
\caption{Initial estimate $\hat {\bf k}_{\up{init}}$ and SML estimate $\hat {\bf k}_{\up{SML}}$, as well as the corresponding relative errors, averaged over the 25 converged Monte-Carlo runs of the third order model.}
\label{table:estimation3}
\begin{tabular}{c|c|c|c|c} 
& $\hat {\bf k}_{\up{init}}$ & ${\boldsymbol \varepsilon}_{\up{init}}$ & $\hat {\bf k}_{\up{SML}}$ & ${\boldsymbol\varepsilon}_{\up{SML}}$ \\ \hline &&&&\\[-3.5mm]
$k_{f1}$ & $4.60\!\cdot\!10^{-17}$ & $2.02\!\cdot\!10^{-2}$ & $4.69\!\cdot\!10^{-17}$ & $1.05\!\cdot\!10^{-5}$ \\
$k_{b1}$ & $9.76\!\cdot\!10^{-28}$ & $4.36\!\cdot\!10^{-3}$ & $9.80\!\cdot\!10^{-28}$ & $4.53\!\cdot\!10^{-6}$ \\
$k_{f2}$ & $4.40\!\cdot\!10^{-20}$ & $6.58\!\cdot\!10^{-4}$ & $4.40\!\cdot\!10^{-20}$ & $4.61\!\cdot\!10^{-7}$ \\
$k_{b2}$ & $2.18\!\cdot\!10^{-21}$ & $4.03\!\cdot\!10^{-3}$ & $2.19\!\cdot\!10^{-21}$ & $2.12\!\cdot\!10^{-6}$ \\
$k_{f3}$ & $1.55\!\cdot\!10^{-16}$ & $6.22\!\cdot\!10^{-3}$ & $1.56\!\cdot\!10^{-16}$ & $1.93\!\cdot\!10^{-6}$ \\
$k_{b3}$ & $2.80\!\cdot\!10^{-28}$ & $1.90\!\cdot\!10^{3}$ & $3.28\!\cdot\!10^{-31}$ & $1.22$ \\
$k_{f4}$ & $1.65\!\cdot\!10^{-19}$ & $1.22\!\cdot\!10^{-1}$ & $1.47\!\cdot\!10^{-19}$ & $9.92\!\cdot\!10^{-6}$ 
\end{tabular}
\end{table}

\begin{figure} [H] 
\centering
\includegraphics[scale = 0.75]{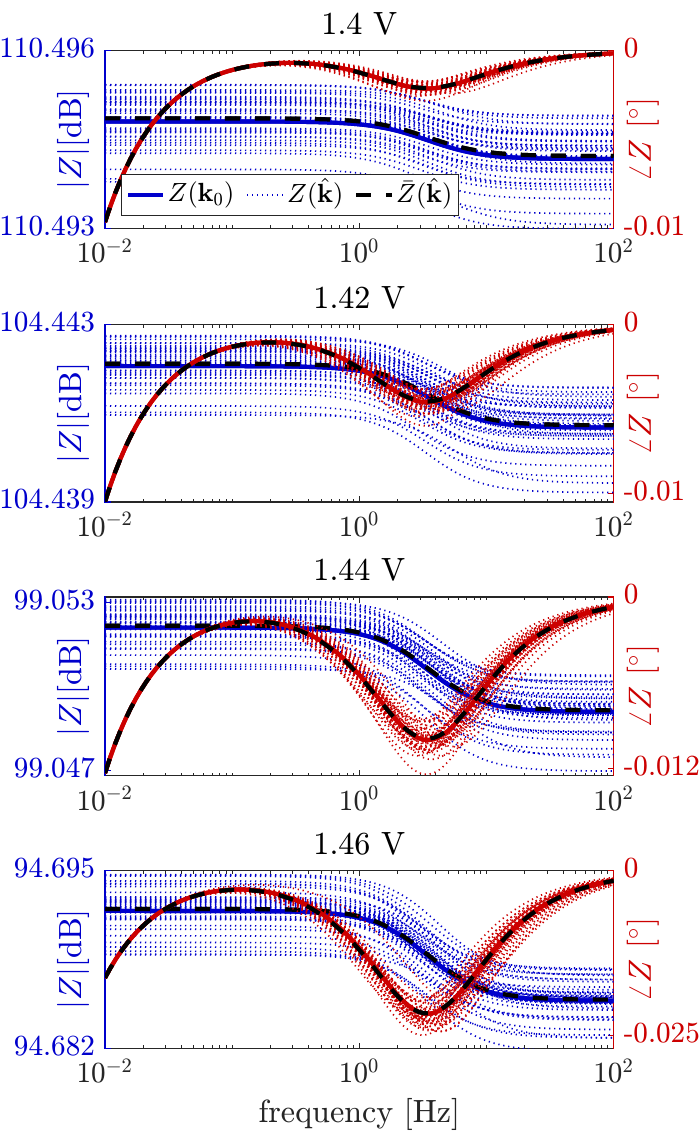} %
\caption{Bode plots of the second order impedance model, evaluated in the true rate constants $Z({\bf k}_0)$ (full coloured line) and in the estimated rate constants $Z(\hat{\bf k})$ (dotted coloured lines) for the 39 converged Monte-Carlo runs. The average estimated impedance $\bar Z(\hat{\bf k})$ is indicated with the black dashed line, and coincides well with the true impedance $Z({\bf k}_0)$.\vspace{-4mm}}
\label{fig:ImpedanceEst2}
\end{figure}

\begin{figure} [H] 
\centering
\includegraphics[scale = 0.75]{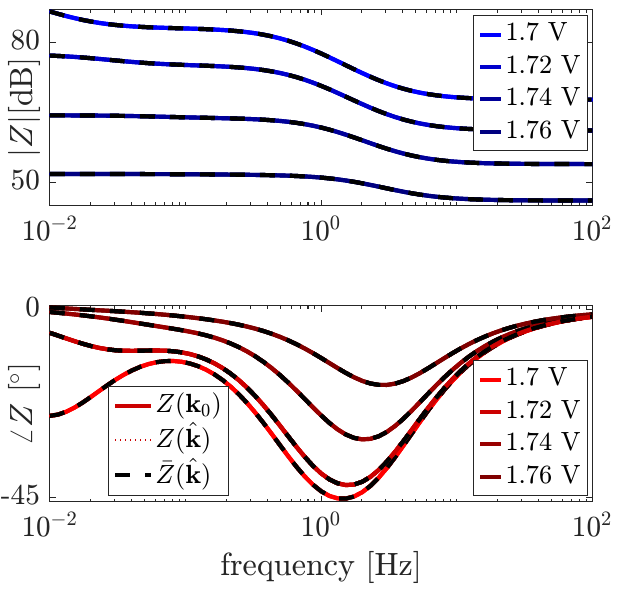} %
\caption{Bode plots of the third order impedance model, evaluated in the true rate constants $Z({\bf k}_0)$ (full coloured line) and in the estimated rate constants $Z(\hat{\bf k})$ (dotted coloured lines) for the 25 converged Monte-Carlo runs. All the estimated impedances $Z(\hat{\bf k})$, and thus also the average estimated impedance $\bar Z(\hat{\bf k})$ (black dashed line), coincide well with the true impedance $Z({\bf k}_0)$.\vspace{-4mm}}
\label{fig:ImpedanceEst3}
\end{figure}

\textbf{Figure}~\ref{fig:ImpedanceEst2} and \textbf{Figure}~\ref{fig:ImpedanceEst3} illustrate how the estimated rate constants $\hat {\bf k}_{\up{SML}}$ can be used to reconstruct the impedance model. 
The dotted lines in Figure~\ref{fig:ImpedanceEst2} show the second order model evaluated in the estimated rate constants for the 39 converged Monte-Carlo runs. 
Except for a small constant factor, the magnitude is generally recovered very well, and also the phase is estimated accurately. 
As a consequence, the average of these estimated impedances $\bar Z(\hat {\bf k})$ corresponds well with the true impedance $Z({\bf k}_0)$. 
For the third order model in Figure~\ref{fig:ImpedanceEst3}, the estimated impedances $Z(\hat {\bf k})$ are actually reconstructed with similar precision as those of the second order model.
However, due to larger dynamic range of the third order impedance, the small deviations with the true impedance $Z({\bf k}_0)$ cannot be visually distinguished.

\subsection{Uncertainty in a given frequency band and over multiple potentials}


As was mentioned in Section~\ref{sec:ID}, 
even if the rate constants are uniquely identifiable from the considered impedance data,
their estimation uncertainty also varies significantly with the chosen frequency band and the number of equilibrium potentials.\\[-1mm]

\begin{figure}[H]
\centering
\includegraphics[scale = 0.75,trim={0 0.7cm 0 0},clip]{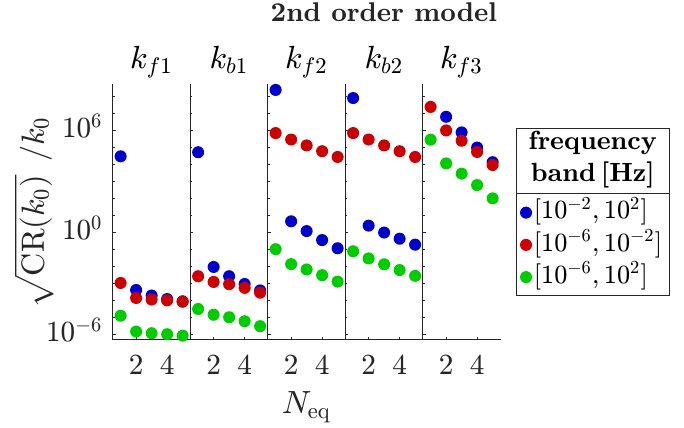} 
\includegraphics[scale = 0.75]{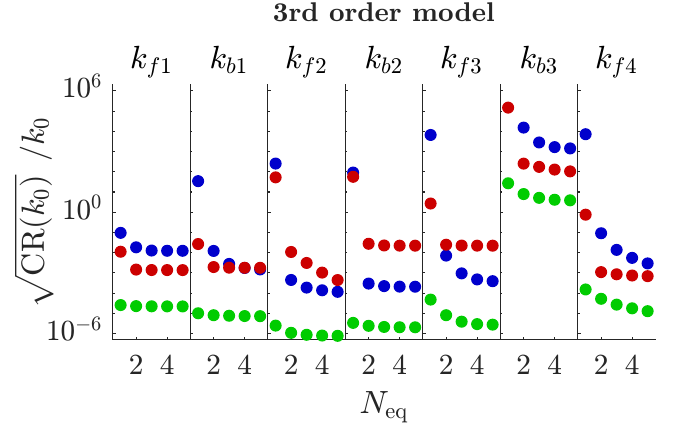} 
\caption{The square root of the Cramér-Rao lower bound on each rate constant, divided by the respective true value of the rate constant, 
expresses the minimum relative uncertainty for a different number of potentials $N_\up{eq}=1,\dots,5$. 
The second order model (top) uses equilibrium potentials between 1.4~V and 1.48~V, while the third order model (bottom) uses potentials between 1.7V and 1.78~V in steps of 0.02~V.  
The colours indicate the chosen frequency band: the upper band $[10^{-2},10^2]$~Hz (blue), the lower band $[10^{-6},10^{-2}]$~Hz (red) and the wide band $[10^{-6},10^2]$~Hz (red).
\vspace{-4mm}}
\label{fig:CRlowerbound}
\end{figure}

\textbf{Figure}~\ref{fig:CRlowerbound} plots the square root of the Cramér-Rao lower bound $\up{CR}(\bf {k}_{0}) = \up{Fi}^{-1}({\bf k}_0)$~\eqref{eq:CR} on each rate constant, scaled by the corresponding true rate constants $\bf {k}_{0}$. 
Remembering that the Cramér-Rao lower bound is the theoretical minimum covariance matrix of the rate constants, this formula can be interpreted as a theoretical minimum of the relative uncertainty on each constant. 
Remark that to calculate the inverse Fisher information matrix \eqref{eq:CR}, 
one needs the noiseless voltage and current DFT spectra ${\bf U}_0 ,{\bf J}_0 \in \mathbb{C}^{N\times1}$,
and the theoretical covariance matrices $C_U = \sigma^2_U I_{N}, C_J = \sigma^2_J I_{N}$ and $ C_{JU} = O_{N\times N}$, where $\sigma_U$ and $\sigma_J$ are calculated from the SNR of 40~dB.\\[-1mm]

Figure~\ref{fig:CRlowerbound} illustrates how the theoretical uncertainty is affected by the frequency band and the number of potentials.
First of all, the plot confirms that the uncertainty decreases, and thus that the informativity increases, if the number of potentials increases.
Especially a single equilibrium potential, which we already established leads to identifiability issues, results in a high uncertainty. 
Moreover, the plot considers three frequency bands in which the current and voltage data is gathered: 
an upper (measurable) band between 10~mHz and 100~Hz, a lower band between 1~$\mu$Hz and 10~mHz, and a wide band that concatenates the other two bands.
In general, the uncertainty decreases if the information of the two bands is combined. 
Whether the upper or the lower band is more informative, depends on the poles and zeros in that band (see Figure~\ref{fig:PathsPolesZeros}). \\[-1mm]



Figure~\ref{fig:CRlowerbound} is consistent with the estimation results in Table~\ref{table:estimation2} and Table~\ref{table:estimation3}.
For the second order model, for example, $k_{f1}$ and $k_{f2}$ are estimated with the smallest relative error, while $k_{f3}$ has the largest relative error. Also for the third order model,  $k_{f2}$ has the smallest and $k_{b3}$ has the largest relative error.  








\subsection{Sensitivity of the cost}


The sensitivity of the maximum likelihood cost to the rate constants can be calculated as
\begin{equation}
   \left(\frac{\partial V_\mathrm{SML}({\bf k})}{\partial {\bf k}}\right)^\top = 2\up{Re}\left(\hat J^H({\bf k}) \hat {\boldsymbol \varepsilon}({\bf k})\right)
\end{equation}
Hence, a large sensitivity to a certain rate constant means that a small variation in the value of that rate constant leads to a large variation in the cost. 
Table~\ref{table:SensitivityCost} shows the relative sensitivity
\begin{equation}
     \left(\frac{\partial V_\mathrm{SML}({\bf k})}{\partial {\bf k}}\right)^\top \circ \,\frac{ {\bf k} }{V_\mathrm{SML}({\bf k})}
\end{equation}
averaged over 100 Monte-Carlo realisations of the disturbing noise, and evaluated in the true rate constants ${\bf k}_0$. 
The results are in line with the observations from the uncertainty analysis, as the rate constants to which the cost is the least sensitive are estimated with the largest uncertainty:  
The cost of the second order model is especially sensitive to $k_{f1}$, but less sensitive to $k_{f3}$. The cost of the third order model cost is the most sensitive to $k_{f2}$, and the least to $k_{b3}$.



\begin{table}[H]
\centering
\caption{Relative sensitivity of the cost with respect to the rate constants}
\label{table:SensitivityCost}
2nd order model\\[1mm]
\begin{tabular}{c|c|c|c} 
 $i$ & 1 & 2 & 3   \\
 \hline&&&\\[-3.5mm]
 $\!\!k_{fi}\!\!$ & $\!\!1.23\!\!\!$  & $\!\!3.31\!\cdot\!10^{-4}\!\!\!$ & $\!\!1.16\!\cdot\!10^{-9}\!\!\!$ \\
 $\!\!k_{bi}\!\!$ & $\!\!3.81\!\cdot\!10^{-1}\!\!\!$ & $\!\!9.46\!\cdot\!10^{-5}\!\!\!$ &  
\end{tabular}\\[3mm]

3rd order model\\[1mm]
\begin{tabular}{c|c|c|c|c} 
 $i$ & 1 & 2 & 3 & 4 \\
 \hline&&&\\[-3.5mm]
 $\!\!k_{fi}\!\!$ & $\!\!9.30\!\cdot\!10^{-5}\!\!\!$  & $\!\!1.47\!\!\!$ & $\!\!2.87\!\cdot\!10^{-1}\!\!\!$ & $\!\!2.74\!\cdot\!10^{-2}\!\!\!$\\
 $\!\!k_{bi}\!\!$ & $\!\!8.21\!\cdot\!10^{-2}\!\!\!$ & $\!\!5.59\!\cdot\!10^{-1}\!\!\!$ & $\!\!1.70\!\cdot\!10^{-7}\!\!\!\!$ & 
\end{tabular}
\end{table}




\section{Conclusion}
The potential-independent reaction rate constants of the multistep OER at a semiconductor photoanode were successfully estimated from simulated electrochemical impedance spectroscopy data at multiple potentials.
First, the fitted impedance model was derived from the microkinetic model, in which the unmeasurable intermediate species coverages were expressed as a function of the rate constants.
It was shown that at lower applied potentials, the order of the model can be reduced. 
For example between 1.55~V and 1.95~V, it suffices to model the OER kinetics by the third order model.

The rate constants were estimated with the sample maximum likelihood estimator, 
integrating not only multiple excited frequencies but also multiple applied potentials.
The SML estimate has the advantage over the traditional least squares estimator of being consistent, i.e. it converges to the true values of the rate constants. 
This consistency is achieved by weighing the minimised equation error with its covariance, which can be calculated by measuring multiple periods of the current and voltage signals.
Due to the large numerical range of the rate constants, different scaling methods were introduced to achieve numerical stability.
To obtain a good initial estimate for the highly nonlinear optimization problem, three different metaheuristic optimization algorithms (ES, GA and PSO) were compared. 
Evolution Strategies appeared to have the highest convergence rate and the lowest average number of cost function evaluations to reach the global minimum.

The rate constants are uniquely identifiable from EIS data if at least two potentials are combined and if the considered frequency band is wide enough to capture the dynamics. Using more than two potentials further decreases the estimation uncertainty. Since some of the dynamics are located at very low frequencies for low potentials and at very high frequencies for high potentials, in practice, only the rate constants from the second and third order impedance model can be estimated accurately.

In this paper, the estimation procedure was validated on noisy, simulated EIS data of PEC with a hematite photoanode.
In the companion paper \cite{Bart2025}, the estimation is applied to actual EIS measurements, which brings some additional challenges. 
First of all, the charge transfer impedance has to be integrated in an equivalent circuit model, whose components have to estimated as well.
Moreover, it is shown that consistency is lost when using impedance data itself, instead of current and voltage data. 












\medskip
\textbf{Acknowledgements} \par 
This work is sponsored by the Strategic Research Program of the Vrije Universiteit Brussel (SRP78). 

\medskip



\bibliographystyle{MSP}
\bibliography{Bibliography.bib}

\end{multicols}





\end{document}